\documentclass[preprint]{aastex63}
\usepackage{amsmath,amssymb}

\begin{document}

\title{Can small-scale magnetic fields be the major cause for the near-surface effect of the solar p-mode frequencies?}

\correspondingauthor{Yan Li}
\email{ly@ynao.ac.cn}

\author{Yan Li}
\affiliation{Yunnan Observatories, Chinese Academy of Sciences, Kunming 650216, China}
\affiliation{Key Laboratory for the Structure and Evolution of Celestial Objects, Chinese Academy of Sciences}
\affiliation{Center for Astronomical Mega-Science, Chinese Academy of Sciences, Beijing 100012, China}
\affiliation{University of Chinese Academy of Sciences, Beijing 100049, China}

\author{Qian-sheng Zhang}
\affiliation{Yunnan Observatories, Chinese Academy of Sciences, Kunming 650216, China}
\affiliation{Key Laboratory for the Structure and Evolution of Celestial Objects, Chinese Academy of Sciences}
\affiliation{Center for Astronomical Mega-Science, Chinese Academy of Sciences, Beijing 100012, China}

\author{Tao Wu}
\affiliation{Yunnan Observatories, Chinese Academy of Sciences, Kunming 650216, China}
\affiliation{Key Laboratory for the Structure and Evolution of Celestial Objects, Chinese Academy of Sciences}
\affiliation{Center for Astronomical Mega-Science, Chinese Academy of Sciences, Beijing 100012, China}

\author{Jie Su}
\affiliation{Yunnan Observatories, Chinese Academy of Sciences, Kunming 650216, China}
\affiliation{Key Laboratory for the Structure and Evolution of Celestial Objects, Chinese Academy of Sciences}
\affiliation{Center for Astronomical Mega-Science, Chinese Academy of Sciences, Beijing 100012, China}

\author{Xing-hao Chen}
\affiliation{Yunnan Observatories, Chinese Academy of Sciences, Kunming 650216, China}
\affiliation{Key Laboratory for the Structure and Evolution of Celestial Objects, Chinese Academy of Sciences}

\author{Gui-fang Lin}
\affiliation{Yunnan Observatories, Chinese Academy of Sciences, Kunming 650216, China}
\affiliation{Key Laboratory for the Structure and Evolution of Celestial Objects, Chinese Academy of Sciences}

\author{Jian-heng Guo}
\affiliation{Yunnan Observatories, Chinese Academy of Sciences, Kunming 650216, China}
\affiliation{Key Laboratory for the Structure and Evolution of Celestial Objects, Chinese Academy of Sciences}
\affiliation{Center for Astronomical Mega-Science, Chinese Academy of Sciences, Beijing 100012, China}
\affiliation{University of Chinese Academy of Sciences, Beijing 100049, China}

\author{Jie-ying Liu}
\affiliation{Yunnan Observatories, Chinese Academy of Sciences, Kunming 650216, China}
\affiliation{Key Laboratory for the Structure and Evolution of Celestial Objects, Chinese Academy of Sciences}

\begin{abstract}
Small-scale magnetic fields are not only the fundamental element of the solar magnetism, but also closely related to the structure of the solar atmosphere. The observations have shown that there is a ubiquitous tangled small-scale magnetic field with a strength of 60 $\sim$ 130\,G in the canopy forming layer of the quiet solar photosphere. On the other hand, the multi-dimensional MHD simulations show that the convective overshooting expels the magnetic field to form the magnetic canopies at a height of about 500\,km in the upper photosphere. However, the distribution of such small-scale ``canopies" in the solar photosphere cannot be rigorously constrained by either observations and numerical simulations. Based on stellar standard models, we identify that these magnetic canopies can act as a global magnetic-arch splicing layer, and find that the reflections of the solar p-mode oscillations at this magnetic-arch splicing layer results in significant improvement on the discrepancy between the observed and calculated p-mode frequencies. The location of the magnetic-arch splicing layer is determined at a height of about 630\,km, and the inferred strength of the magnetic field is about 90\,G. These features of the magnetic-arch splicing layer derived independently in the present study are quantitatively in agreement with the presence of small-scale magnetic canopies as those obtained by the observations and 3-D MHD simulations.
\end{abstract}


\section{Introduction}

In 1962, Leighton and co-workers discovered rich oscillations with periods of approximately 5 minutes on the surface of the sun (Leighton, Noyes, and Simon 1962). Later in 1975, Deubner (1975) confirmed that these oscillations are global resonant waves with pressure as the main restoring force, which are now named as the solar p-mode oscillations. Frequencies of millions of solar p-mode oscillations have been measured by grand-based networks, for example GONG (Harvey et al. 1996), and spaced-based telescopes, for example SOHO (Schou et al. 1997) with extremely high precisions, and have been used to extract information of the solar interior.

The internal structure of the sun is described by the standard solar models. They are based on the up-to-date physics, and computed with state-of-the-art stellar evolution codes, for example the Modules for Experiments in Stellar Astrophysics (Paxton et al. 2011, 2013, 2015, 2018; referred to hereafter as MESA). The standard solar models can reproduce, for example, the distribution of the sound speed in the solar interior that is in agreement with the one derived from helioseismic inversions with fairly high precision (Christensen-Dalsgaard et al. 1985). For frequencies of individual solar p-mode oscillations, however, the standard solar models usually produce theoretical oscillation frequencies systematically offset from the observed ones. This is the so-called near-surface effect of solar p-mode oscillations (Christensen-Dalsgaard, D\"appen, and Lebreton 1988; Dziembowski, Paterno, and Ventura 1988), whose physical nature has not been fully understood up to now.

In order to eliminate the near-surface effect on the applications of the observed frequencies in explorations of solar and stellar interior, various fitting formulae have been proposed to correct the regular part of the frequency offset between observations and model calculations. Kjeldsen, Bedding, and Christensen-Dalsgaard (2008) proposed a power-law correction, which takes the form of $\delta\nu = a\nu^b $ where $\nu$ is the frequency and $\delta\nu$ is the frequency correction. According to the solar calibrations, the parameter $b = 4.9$ has been adopted to correct frequencies of pulsating stars other than the Sun (e.g., ChristensenDalsgaard et al. 2010). Ball and Gizon (2014) suggested two formulae: one is $\delta\nu = a_3\nu^3/I $ and the other is $\delta\nu = (a_{-1}\nu^{-1} + a_3\nu^3)/I $, where $I$ is the normalized mode inertia. On the other hand, the structure of the sun in the vicinity of its photosphere has carefully been examined by nonstandard solar models (Demarque, Guenther, and Kim 1997; Rosenthal et al. 1999) and by hydrodynamic simulations of 1D, 2D, and 3D configurations (Carlsson and Stein 1994; Leenaarts et al. 2007; Wedemeyer et al. 2004). Recently, modifications due to interactions between p-mode oscillations and turbulent convection near the top of the solar convection zone were considered as the possible cause of the near-surface effect (e.g., Ball et al. 2016; Houdek et al. 2017; Schou \& Birch 2020).

Small-scale magnetic fields of the quiet solar photosphere are an important component of the solar magnetism. Because of their small sizes on the solar disk, they cannot be seen on the solar magnetograms and are sometimes referred to as ``hidden magnetic fields"(Stenflo 1987, 2013). Based on 3-D realistic hydrodynamical simulations of solar surface convection, Trujillio Bueno et al. (2004) derived an averaged un-signed flux density of at least 60\,G with the aid of the observed Hanle depolarization. With the spatial resolution of 0.3$''$, the Solar Optical Telescope onboard the Hinode space observatory allows for unprecedented polarimetric observations of the magnetic field of the quiet sun (Kosugi et al. 2007). The observations show vertical field components with kilo-Gauss strength and concentrated in the intergranular lanes, along with horizontal field components with ten-Gauss strength and occurring at the edges of bright granules. Lites et al. (2008) determined for the horizontal field component a mean apparent field strength of 55\,G, while for the vertical field component a corresponding mean absolute strength of only 11\,G. Unlike the magnetic fields of the sun spots that are isolated on the solar disk and occupy only a very small fraction of the solar surface, the small-scale magnetic fields are ubiquitous on the solar surface and thus store up a great amount of magnetic energy, which can be an important source for the magnetic coupling to the outer atmosphere and the corona heating (Stenflo 1994; Schrijver et al. 1998; Schrijver \& Title 2003). Furthermore, such ``turbulent" magnetic fields can significantly affect the structure of the solar atmosphere and the propagations of acoustic waves (Rosenthal et al. 2002; Bogdan et al. 2003; Cally 2007).

More observational studies have been made in recent years to probe the internetwork magnetic fields. Danilovic et al. (2010) made observations of the transverse component of the photospheric magnetic fields by use of the magnetograph SUNRISE/IMAX. Using the infrared spectropolarimeter GRIS equipped on the GREGOR telescope, Lagg et al. (2016) made highly sensitive full Stokes measurements in the 1.56$\mu$m Fe\,I lines and found that about 80\% of the obtained spectra show polarimetric signals. With more detailed analyses, Mart\'inez Gonz\'alez et al. (2016) found that half of the observed region can be explained by the magnetic fields with spatial scales smaller than the resolution element. Two magnetic field populations are distinguished: one with the larger filling factor has weak horizontal fields of about $\sim$150\,G, the other with the smaller filling factor is constrained for the field strengths above 250\,G. Using the high-spatial resolution full-Stokes measurements obtained by the CRISP equipped on the SST, Morosin et al. (2020) derived the depth-stratification of the magnetic field in a plage region. The magnetic fields emanate from the intergranular lanes in the photosphere and spread out horizontally toward the chromosphere to form a canopy. Based on similar data, Keys et al. (2020) found evidence for fast amplification of magnetic field strength in the magnetic bright points during their temporal evolution to kilogauss field strength. Observations of resonance scattering polarization with high sensitivity and high spatial resolution are also available to diagnose small-scale dynamos and magnetoconvection (Zeuner et al. 2018, 2020). 

More information about the small-scale magnetic fields is provided by the multi-dimensional magneto-hydrodynamic (MHD) simulations. In the 2-D MHD simulations, Grossmann-Doerth et al. (1998) found that the magnetic flux are quickly swept by the horizontal convective flows into the intergranular downflow channels, while strong horizontal magnetic fields are pushed upward by the convective upflows. Horizontal magnetic fields were frequently found in the 3-D MHD simulations of Schaffenberger et al. (2005, 2006), and often referred to as ``small-scale canopies". The 3-D MHD simulations of Abbett (2007) showed horizontal ribbons of the magnetic flux with magnetic voids located in the below. Sch\"ussler \& V\"ogler (2008) and Rempel (2014) demonstrated in their radiative MHD simulations that a local dynamo can operate in the solar photosphere, which maintains a turbulent distribution of small-scale magnetic fields. In simulations of Steiner et al. (2008), a local maximum is found for the horizontal field component in the upper photosphere, whose values can surpass that of the corresponding vertical component by a factor of 2 to 5.6 depending on the different choices of initial states and boundary conditions. Coincidentally, Rempel (2014) also found that the ratio of the horizontal to vertical field shows a maximum at similar height as Steiner et al., with its value from about 3 to 5 depending on the overall field strength decreasing from 60\,G to 20\,G. 

The coupling between the small-scale magnetic fields and the acoustic waves mainly has two aspects. The small-scale magnetic fields modify the hydrostatic structure of the atmosphere, and thus can significantly affect the propagation of the acoustic waves. On the other hand, the magnetic field can act as an additional restoring force, and then gradually change the waves from purely acoustic to magneto-acoustic. Rosenthal et al. (2002) pointed out that due to rapidly increasing phase speed of the fast magneto-acoustic modes, waves will be refracted along a magnetic field that is significantly inclined to the vertical direction, resulting in finally total internal reflection in the atmosphere. Using the phase lag of the Doppler signal in different layers in the photosphere and low chromosphere, Finsterle et al. (2004) determined the 3-D topography of the magnetic canopy in and around active regions. Steiner et al. (2007) found from the direct numerical simulations that a plane-parallel acoustic wave is partially refracted and reflected by the interaction with the magnetic field. They pointed out further that the perturbation of the wave front is not due to the presence of a magnetic field, but rather to the temperature decreasing associated with the vigorous intergranular downflows.

The accumulation of the horizontal magnetic flux is considered as a direct result of the convective overshooting beyond the solar convection zone, which expels the horizontal component of the magnetic field into the upper photosphere. Because of similar sizes of the individual granules, the resulted magnetic arches will be formed at a similar height and may be spliced as a global layer which we refer to hereafter as the magnetic-arch splicing layer. In this work, we identify the existence of such a magnetic-arch splicing layer in the solar atmosphere, and find that the inclusion of the reflections of the solar p-mode oscillations on the magnetic-arch splicing layer significantly improves the agreement of the theoretical solar p-mode frequencies with the observed ones. In Section 2, we briefly introduce the input physics and methods in our calculations of stellar models. We discuss our main results in Section 3, and summarize our main conclusions in Section 4.

\section{Input physics and methods of calculations}

Frequencies of solar p-mode oscillations are taken from observations of the Birmingham Solar-Oscillation Network (BiSON, see Broomhall et al. 2009). The observations ran at 6 stations distributed all around the world and lasted 8640 days from 1985 to 2008. Frequencies of low spherical harmonic degrees ($l$ = 0, 1, 2, and 3) are estimated to unprecedented levels of precision. We use these p-mode frequencies in the fitting of stellar models to the sun.

The stellar models are constructed by the MESA. The specific version we have adopted is version 11554, which incorporates the up-to-date input physics in calculations of stellar structure and evolution. In particular, we adopt the OPAL equation of state tables (Rogers \& Nayfonov 2002). The OPAL opacities (Iglesias \& Rogers 1996) are used in the high temperature region, while the opacities from Ferguson et al.(2005) are used in the low temperature region to include contributions from molecules and grains. We use the classical mixing-length theory (MLT) to treat the convection. In order to satisfy several observational restrictions (such as the depth of the convection zone, the surface helium abundance, etc.), we adopt GS98 chemical composition (Grevesse \& Sauval 1998) and include element diffusion from Thoul et al. (1994) in our solar models. Frequencies of the p-mode oscillations for the considered stellar models are computed with ``adipls" package in MESA, which is developed and updated by Christensen-Dalsgaard (2008).

In order to approximately include the effects of convection and overshooting in the the solar atmosphere, we adopt two different $T$-$\tau$ relations in the atmosphere integration. One is the Krishna Swamy relation (referred to as the K-S atmosphere, see Krishna Swamy 1966), and the other is the fit given by Sonoi et al. (2019) to the Vernazza et al. VAL-C model (Vernazza et al. 1981, referred to as the Hopf-grey atmosphere). As pointed out by Sonoi et al. (2019), different $T$-$\tau$ relations can significantly affect the entropy profile in the solar atmosphere, and the VAL-C model actually gives satisfactory correspondence to the results of the 3-D radiation-coupled hydrodynamical simulations. The two $T$-$\tau$ relations can be written in a unified form:
\begin{equation}\label{1}
{{T}^{4}}=\frac{3}{4}T_{\text{eff}}^{4}\left[ \frac{}{}\tau + q(\tau ) \right],
\end{equation}
where $T$ is temperature, $\tau$ is the optical depth, and $q(\tau )$ is a Hopf-like function. $T_{\text{eff}}$ is the effective temperature of the star. It should be noted that we define the base of the photosphere at an optical depth of $\tau=0.412$ for the solar-Hopf atmosphere (Sonoi et al. 2019) and $\tau=0.312$ for the Krishna Swamy atmosphere (Krishna Swamy 1966), while in the 3-D simulations it is defined at about $\tau=1$. These different height systems in the solar photosphere models may introduce some difference in the results of MHD simulations and stellar standard models.

We introduce the magnetic field in the solar atmosphere by adding an additional term in the Hopf-like function:
\begin{equation}\label{2}
q(\tau )={{q}_{\text{ori}}}(\tau )+a\exp \left( -b\sqrt{\tau } \right),
\end{equation}
where ${{q}_{\text{ori}}}(\tau )$ is the original Hopf-like function that is given by the two considered $T$-$\tau$ relations, respectively. This additional term results in an increase of temperature in the upper photosphere, and the corresponding increase of the radiative pressure takes the role of the magnetic pressure. $a$ and $b$ are two parameters to be determined for each specific $T$-$\tau$ relations. In particular, parameter $a$ mainly determines the strength of the magnetic field, while parameter $b$ specifies the location of the magnetic-arch splicing layer. Table 1 summaries our main results. In addition to the guess function of square root in Eq.\,(2), we have tested a few more guess functions (including linear, square, and cubic), and the results are similar with only small variations on $\chi^2$ (around ten percent), which is defined as the sum of the squared differences of frequencies between observed and calculated ones weighted by the observational errors. Therefore we adopt the function of square root that minimizes the $\chi^2$ among the other functions we have tested.

\begin{table*}
\caption{Properties of the best-fit solar models}
\centering
\begin{tabular}{cccccccccccc}
\hline
\hline
$\text{[Fe/H]}_{\text{ini}}$ & $Z_{\text{ini}}$ & $Y_{\text{ini}}$ &
$\alpha$ & $R/R_{\odot}$ & $\bigtriangleup L/L_{\odot}$ &
$Y_{\text{s}}$ & $\left( Z/X \right)_{\text{s}}$ & 
$R_{\text{CZ}}/R_{\odot}$ & $\chi^2$ \\
\hline
\multicolumn2c{} & \multicolumn3c{K-S atmosphere} & 
\multicolumn3c{$300\exp \left( -98\sqrt{\tau } \right)$} & \multicolumn2c{} \\
\hline
0.08 & 0.01941 & 0.27634 & 2.11610 & 0.999527 & 0.000024 & 0.24899 & 0.02429 & 0.71819 & 240.0  \\
0.09 & 0.01979 & 0.27849 & 2.12535 & 0.999614 & 0.000085 & 0.25116 & 0.02488 & 0.71617 & 219.5  \\
0.10 & 0.02019 & 0.28046 & 2.13595 & 0.999723 & 0.000047 & 0.25318 & 0.02548 & 0.71523 & 212.0  \\
0.11 & 0.02060 & 0.28206 & 2.14650 & 0.999800 & 0.000063 & 0.25481 & 0.02608 & 0.71364 & 220.2  \\
0.12 & 0.02102 & 0.28360 & 2.15490 & 0.999841 & 0.000020 & 0.25636 & 0.02670 & 0.71282 & 243.5  \\
\hline
\multicolumn2c{} & \multicolumn3c{Hopf-grey atmosphere} & 
\multicolumn3c{$310\exp \left( -86\sqrt{\tau } \right)$} & \multicolumn2c{} \\
\hline
0.08 & 0.01941 & 0.27634 & 2.02345 & 0.999489 & 0.000024 & 0.24899 & 0.02429 & 0.71736 & 231.0  \\
0.09 & 0.01979 & 0.27849 & 2.03270 & 0.999579 & 0.000086 & 0.25116 & 0.02488 & 0.71694 & 215.2  \\
0.10 & 0.02019 & 0.28046 & 2.04327 & 0.999677 & 0.000032 & 0.25318 & 0.02548 & 0.71485 & 213.4  \\
0.11 & 0.02060 & 0.28206 & 2.05387 & 0.999763 & 0.000063 & 0.25481 & 0.02608 & 0.71417 & 226.8  \\
0.12 & 0.02102 & 0.28360 & 2.06238 & 0.999801 & 0.000010 & 0.25636 & 0.02670 & 0.71281 & 252.0  \\
\hline
\end{tabular}

\tablecomments{${\rm [Fe/H]}_{\rm ini}$ is the initial metallicity,
  $Z_{\rm ini}$ is the initial metal abundance,
  $Y_{\rm ini}$ is the initial helium abundance,
  $\alpha$ is the mixing-length parameter which is the ratio of 
           the mixing-length to the local pressure scale height,
  $R/R_{\odot}$ is the relative radius,
  $\bigtriangleup L/L_{\odot}$ is the relative difference of luminosity,
  $Y_{\rm s}$ is the helium abundance at the stellar surface,
  $(Z/X)_{\rm surf}$ is the ratio of the metal abundance to the hydrogen 
           abundance at the stellar surface,
  $R_{\rm CZ}/R_{\odot}$ is the depth of the convection zone,
  and $\chi^2$ is the sum of the squared differences of frequencies between 
  observed and calculated ones weighted by the observational errors.} 

\end{table*}

An important issue that is related to the reflection and the transmission of the p-mode oscillations on the magnetic-arch splicing layer is the suitable form for one of the surface boundary conditions in the calculations of the linear adiabatic oscillations, i.e., the mechanical surface boundary condition. In an isothermal atmosphere without the magnetic field, for example the Eddington gray atmosphere, the mechanical surface boundary condition of the linear adiabatic oscillations is given by Unno et al. (1989)
\begin{equation}\label{3}
\left( 1+\frac{{{H}_{P}}}{r}\left[ \frac{l\left( l+1 \right)g}{{{\omega }^{2}}r}-4-{{\omega }^{2}}\frac{r}{g} \right] \right){{\xi }_{r}}-\frac{1}{\rho g}P'=0,
\end{equation}
where ${\xi }_{r}$ is the radial displacement, $P'$ is the Eulerian perturbation of pressure, $\rho$ is density, $g$ is the gravitational acceleration, $\omega$ is the circular frequency of the pulsation mode, and $H_P$ is the local pressure scale height at radius $r$. For those p-modes with frequencies that are higher than the acoustic cut-off frequency, the reflection condition on a free boundary is usually adopted:
\begin{equation}\label{4}
\delta P=P'-\rho g{{\xi }_{r}}=0,
\end{equation}
where $\delta P$ is the Lagrangian perturbation of pressure (for more discussions, see Unno et al. 1989; Christensen-Dalsgaard 2008). With the inclusion of the magnetic field in the atmosphere, the magnetic pressure increases rapidly in the upper photosphere, resulting in a rapid decrease of density as well as a rapid increase of the sound speed, and thus Eqs.\,(3) and (4) no longer hold. Therefore, we need to use a new mechanical surface boundary condition. For this aim, it is instructive to consider the propagations of the p-mode oscillations in the upper photosphere. When the acoustic waves propagate outwardly and finally meet the magnetic-arch splicing layer, they will be reflected and transmitted. For those waves that are totally reflected at the magnetic-arch splicing layer, standing waves will be formed for the specific eigen-frequencies. Those acoustic waves are exactly what we refer to as the solar p-mode oscillations. Other waves that can transmit into the solar chromosphere cannot form standing waves, and will not be observed as global oscillation modes. As the solar p-mode oscillations should be totally reflected at the magnetic-arch splicing layer, there is no wave beyond the magnetic-arch splicing layer, and a suitable surface mechanical boundary condition is
\begin{equation}\label{5}
P'=0.
\end{equation}
The two surface boundary conditions, i.e. Eqs.\,(4) and (5), are all applied in our calculations of the solar p-mode oscillations, and comparisons and discussions are made to illustrate their effects on the frequencies of the p-mode oscillations.

\section{ Results and discussions }

The observed solar p-mode oscillations are eigen-oscillation modes of the sun with the pressure as the main restoring force. They propagate in the solar interior and atmosphere, taking the whole sun as a huge resonant cavity. In order to reproduce the observed frequencies, we need to know the internal structure of the sun. The so-called standard solar models are based on the up-to-date physics. By adjusting usually the initial helium abundance and the mixing-length parameter, a stellar model with exactly one solar mass ($M_{\odot}=1.9892\times 10^{33}$\,g) is calibrated to have the radius and the luminosity of the present sun ($R_{\odot}=6.9598\times 10^{10}$\,cm and $L_{\odot}=3.8418\times 10^{33}$\,erg\,s$^{-1}$) after evolving from the zero-age main sequence to the present solar age (4.57Gyr). Based on the standard solar models, the frequencies of the eigen-oscillation modes are calculated with the theory of the linear adiabatic oscillations. Compared with the observed frequencies, we have a good chance to check the theory of stellar structure and evolution (Bahcall et al. 2005; Basu 2016). However, the standard solar models always result in the theoretical oscillation frequencies that are systematically offset from the observed ones. This is the well-known ``surface term" problem (see, e.g., Basu 2016), whose physical nature has not yet been fully understood up to know. 

In order to study reflection and transmission of the solar p-mode oscillations at the magnetic-arch splicing layer in the upper photosphere, we approximate the effect of the magnetic field by adding in the solar atmosphere model a magnetic pressure with two parameters, one determining the location of the magnetic-arch splicing layer and the other specifying the strength of the magnetic field. The total pressure is thus composed of the gas pressure and the magnetic pressure. The increase of the magnetic pressure will result in the decrease of the gas pressure due to the hydrostatic equilibrium of the atmosphere, and further the decrease of the density which leads to the reflections of the solar p-mode oscillations due to the rapid increase of the sound speed. By adjusting these two parameters of the magnetic field we find out the best-fit models for the observed solar p-mode oscillations. 

\begin{figure}
\plotone{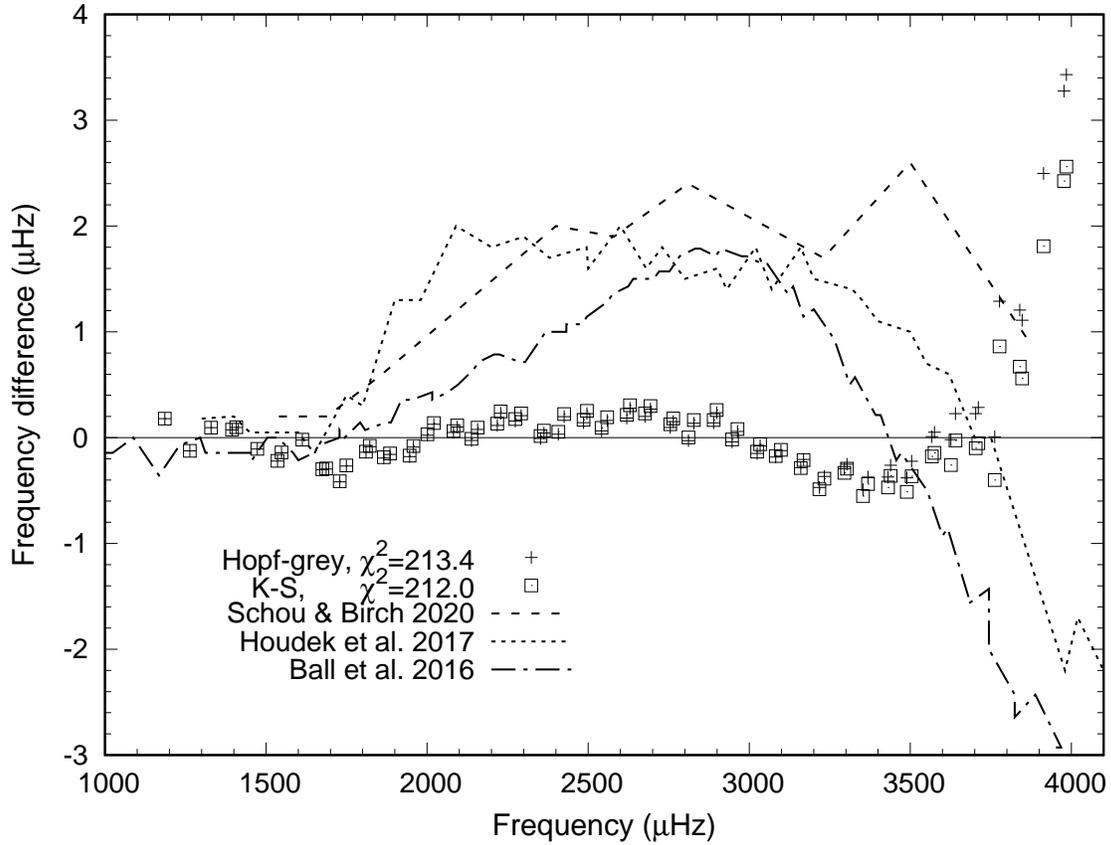}
\figurenum{1}
\caption{The frequency differences between observations and model calculations of the best-fit solar models with different solar atmosphere models. The frequencies of the solar p-mode oscillations are taken from BiSON observations (Broomhall et al. 2009). Results of some recent investigations (Ball et al. 2016; Houdek et al. 2017; Schou \& Birch 2020) are also shown for comparison.} 
\end{figure}

In the present study, we do not follow  the usual procedure to do the calibration of the standard solar models. But instead, we only match the stellar models to the present solar luminosity. For the stellar models near the solar age in each evolutionary sequence, we compare the frequencies of their p-mode oscillations with the observed values of the corresponding modes of the sun, and choose the best-fit model by the least of $\chi^2$ that is defined as the sum of the squared differences of frequencies between observed and calculated ones weighted by the observational errors. We then adjust values of the mixing-length parameter and the initial helium abundance iteratively to make the best-fit model being exactly of the solar age. The differences between the observed frequencies and the corresponding model calculated frequencies are shown for the best-fit solar models with two different atmosphere models in Figure 1, i.e., plus sign for the results with the Hopf-grey atmosphere and the dotted-square sign for the results with the K-S atmosphere. For comparison, the similar results of some other recent investigations (e.g., Ball et al. 2016; Houdek et al. 2017; Schou \& Birch 2020) are also shown in the same figure, with different approaches to treat the interactions between p-mode oscillations and turbulent convection. The turbulent convection was simulated by a 3-D radiation hydrodynamics code MURaM in Ball et al. (2016) and Schou \& Birch (2020). In particular, Ball et al. (2016) used MESA to calculate the standard solar model and then replaced its near-surface equilibrium structure with the results of MURaM. The p-mode frequencies were calculated and compared for the stellar models with and without the improved near-surface equilibrium structure. In order to evaluate the frequency perturbations resulted from the dynamical part of the interaction of the oscillations with near-surface convection, Schou \& Birch (2020) numerically estimated eigenfunctions in the 3-D hydrodynamic simulations and matched them to the eigenfunctions calculated from the classic equations applied to the horizontally averaged variables of the stellar structure. Houdek et al. (2017) also adopted the approach of replacing the outer layers of a 1-D solar envelope model by the structure of an averaged 3-D simulation, and moreover they include the effects of non-adiabaticity and convection dynamics by using a 1-D non-local and time-dependent convection model. Compared our results with those of other investigations, significant improvement is achieved on the discrepancy between calculated and observed solar p-mode frequencies. It should be noticed that the original systematic offset (a power-law type, e.g. Kjeldsen et al. 2008) between the observed and the calculated p-mode frequencies is completely eliminated. The residuals now left behind in our results are still larger than the observational errors. It can be seen that the residuals show an approximate sinusoidal-type distribution when the frequency is smaller than about 3700$\mu$Hz, and then rapidly increase. As pointed out by Gough (1990), a localized feature in the sound speed inside a star introduces an oscillatory term in the frequencies of the p-modes as a function of frequency. Based on this conclusion, we can infer for the sun that the localized feature indicated by the present sinusoidal-type residuals is in the deep solar convection zone below the solar photosphere, where the turbulent pressure may contribute a considerable fraction to the local hydrostatic equilibrium. Finally, it can be noticed that only a few p-modes with the highest frequencies for each spherical harmonic degree ($l=0$, 1, 2, and 3) have considerably higher frequencies than the corresponding calculated ones, which is seen as the steep increase for the frequency above 3700\,$\mu$Hz in Figure 1. Without the magnetic-arch splicing layer, however, the observed frequencies of those p-modes are significantly lower than the calculated ones as seen in Figure 4 of Basu (2016). These facts indicate that those p-modes do not totally reflected at the magnetic-arch splicing layer and a small fraction of the acoustic waves transmits into the upper atmosphere. For other p-mode oscillations with even higher frequencies, most of the acoustic waves will transmit into the upper atmosphere and they cannot be observed as the eigen-modes of the sun. In particular, we estimate the acoustic cut-off frequency in the photosphere by 
\begin{equation}\label{6}
{{\omega }_{a}}=\frac{{{c}_{s}}}{2{{H}_{p}}}\text{=}\sqrt{{{\Gamma }_{1}}\frac{\mu }{\Re T}}\frac{g}{2} ,
\end{equation}
which is defined in an isothermal atmosphere. In Eq.\,(6), $c_s$ is the sound speed, $H_p$ is the local pressure scale height, $\Gamma_1$ is the adiabatic exponent, $\mu$ is the mean molecular weight, $\Re$ is the gas constant, and $g$ is the gravity acceleration. The result is about 6400$\mu$Hz, that is in agreement with other results in the literature. This value is much higher than the frequencies of the considered p-modes, and is not in agreement with the results in our work (around 3700$\mu$Hz). The reason most probably comes from the isothermal atmosphere assumption, which is not valid in our solar models. Due to this reason, it is difficult for us to theoretically determine the exact cut-off frequency for the considered p-mode oscillations.

\begin{figure}
\plotone{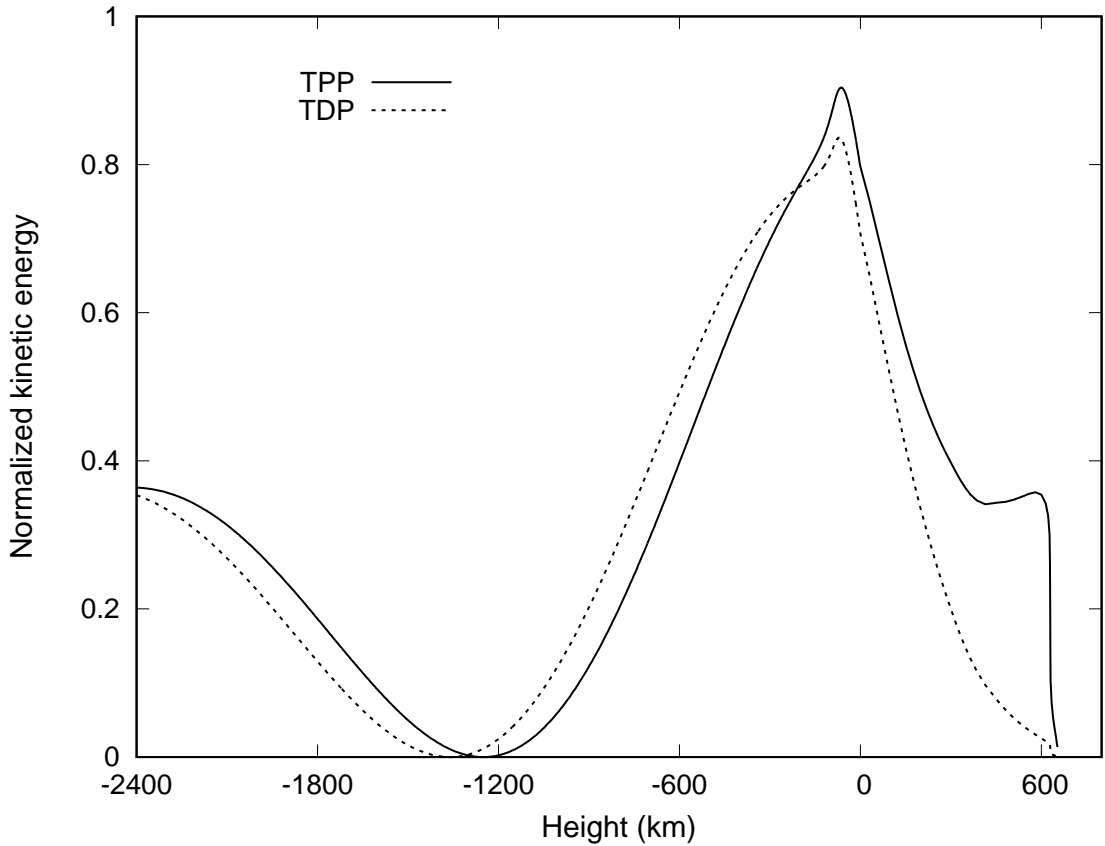}
\figurenum{2}
\caption{The distributions of the normalized kinetic energy by adopting two different mechanical surface boundary conditions are shown for a particular $l=1$ mode corresponding to the observed frequency of 3233.14$\mu$Hz of the solar model with the inclusion of the magnetic-arch splicing layer. The solid line represents the one by adopting Eq.\,(5) (TPP model), while the dotted line by adopting Eq.\,(4) (TDP model) at the surface of the atmosphere. The two lines are matched with each other in the stellar interior, to highlight their differences in the atmosphere.}
\end{figure}

\begin{figure}
\plotone{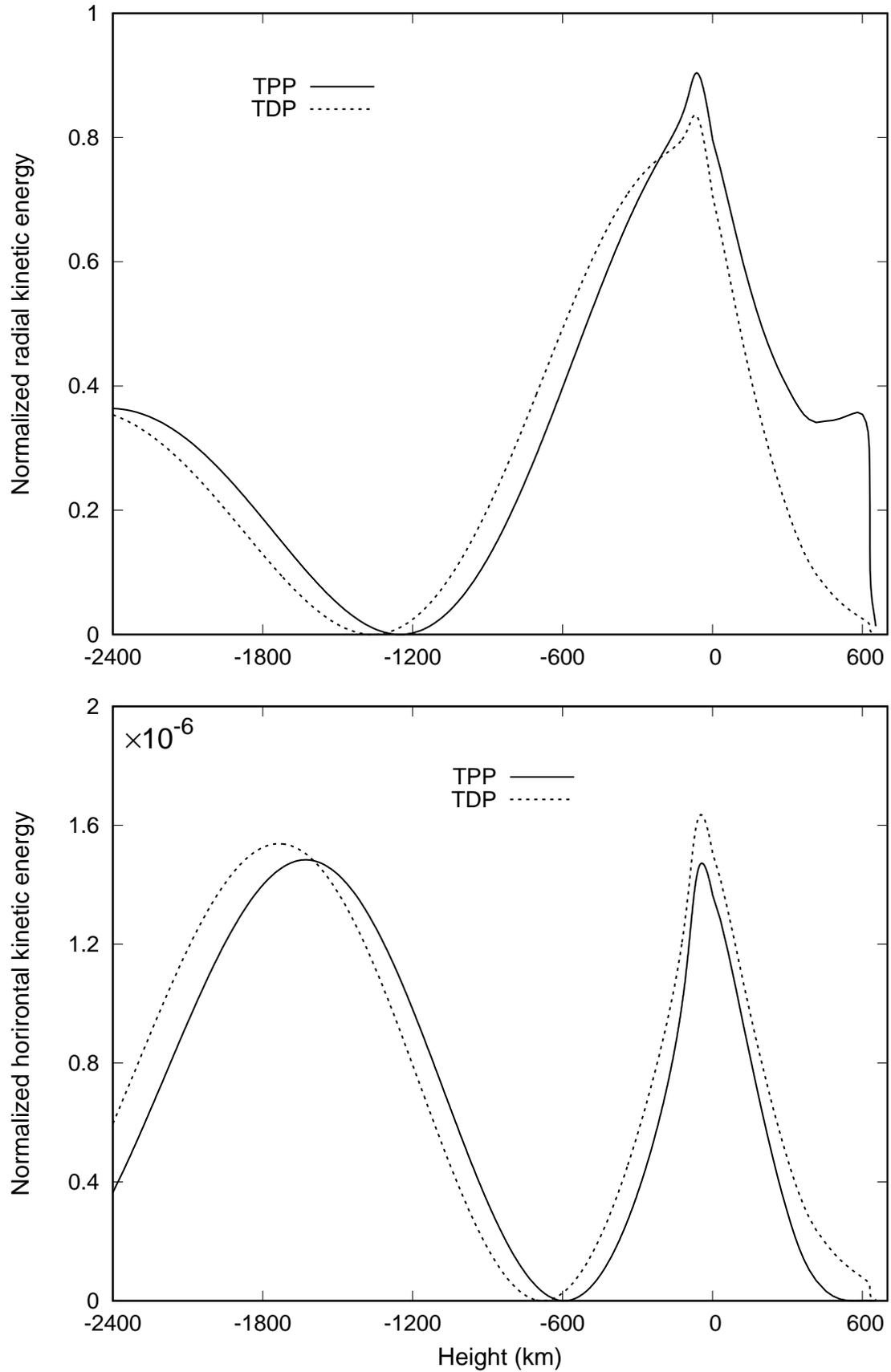}
\figurenum{3}
\caption{The distributions of the normalized radial kinetic energy are shown in the upper panel while the distributions of the normalized horizontal kinetic energy are shown in the lower panel for the same oscillation mode as in Figure 2. In the lower panel, numbers on the ordinate should be multiplied by $10^{-6}$ to obtain the same unit as in the upper panel.}
\end{figure}

  We illustrate why the reflections of acoustic waves on the magnetic-arch splicing layer can significantly improve the agreement between observed and calculated frequencies of the solar p-mode oscillation by examining the propagations of p-mode oscillations near the magnetic-arch splicing layer. We consider the distribution of the normalized kinetic energy of an oscillation mode that is defined by 
\begin{equation}\label{7}
{{e}_{k}}=\left[ \xi _{r}^{2}+l(l+1)\xi _{h}^{2} \right]\rho {{r}^{2}},
\end{equation}
where $\xi _{h}$ is the horizontal displacement (see Unno et al. 1989; Aerts et al. 2010 for more details). Figure 2 shows the distributions of the normalized kinetic energy for a particular $l=1$ mode with the observed frequency of 3233.14$\mu$Hz, computed with different surface mechanical boundary conditions Eq.\,(4) (which will be referred to as TDP model) and Eq.\,(5) (which will be referred to as TPP model). We obtain the radial and horizontal displacement of a p-mode oscillation by solving the equations of the linear adiabatic oscillation (see, e.g., Christensen-Dalsgaard 2008). According to the theory of linear adiabatic oscillation, the solutions can be multiplied by any constant, and the usual procedure is to set the normalization condition at the outer boundary of the considered stellar structure (see Christensen-Dalsgaard 2008 for more details). As a result, different outer boundary conditions will lead to different values of the radial and horizontal displacement in the stellar interior. In order to compare the radial and horizontal displacement under different outer boundary conditions, we match the two distributions of the normalized kinetic energy in the solar interior. Therefore Figure 2 clearly shows their difference near the magnetic-arch splicing layer. It can be seen that there is a bump just below the magnetic-arch splicing layer for the TPP model, which is not seen for the TDP model and for the other cases that the magnetic-arch splicing layer is not included in the solar models. Such a bump in the normalized kinetic energy equivalently enlarges the propagation region of the acoustic waves, which can be recognized easily by comparing the positions of two corresponding nodes. A larger propagation region makes the p-mode oscillations to spend a little longer times to travel across and therefore to have a little lower frequencies.

The outer boundary condition is crucial for the reflection of the p-mode oscillations at the magnetic-arch splicing layer. We show this by analyzing the distribution of the pressure perturbation in this near-surface stellar structure. In Figure 3 we show the distributions of radial kinetic energy, which is defined by the first term in the right-hand-side of Eq.\,(7), and of horizontal kinetic energy which is defined by the second term of Eq.\,(7) for the same oscillation mode as in Figure 2. It can be noticed from the lower panel of Figure 3 that the horizontal kinetic energy of the TPP model is extremely small around the magnetic-arch splicing layer. As the horizontal kinetic energy is directly related to the Eulerian pressure perturbation, there is actually no acoustic wave above the magnetic-arch splicing layer with the use of the surface boundary condition Eq.\,(5). As a result, we can see in the upper panel of Figure 3 that the distribution of the radial kinetic energy is flat below the magnetic-arch splicing layer, indicating essentially total reflections of acoustic waves on a node of the Eulerian pressure perturbation. With the use of the surface boundary condition Eq.\,(4), however, the Eulerian pressure perturbation is not zero at the outer boundary of the near-surface stellar structure due to the normalization condition that demands the radial displacement to be a finite value at the outer boundary. This is equivalent to an input of acoustic waves from outside into the stellar interior. As a result, we can see in the lower panel of Figure 3 that a considerable amount of the horizontal kinetic energy is present above the magnetic-arch splicing layer for the TDP model. Consequently, we can see in the upper panel of Figure 3 that the radial kinetic energy decreases rapidly after going outward across the magnetic-arch splicing layer. This effect considerably reduces the cavity of the p-mode oscillations, and makes their frequencies higher than the corresponding ones calculated with the surface boundary condition Eq.\,(5).

\begin{figure}
\plotone{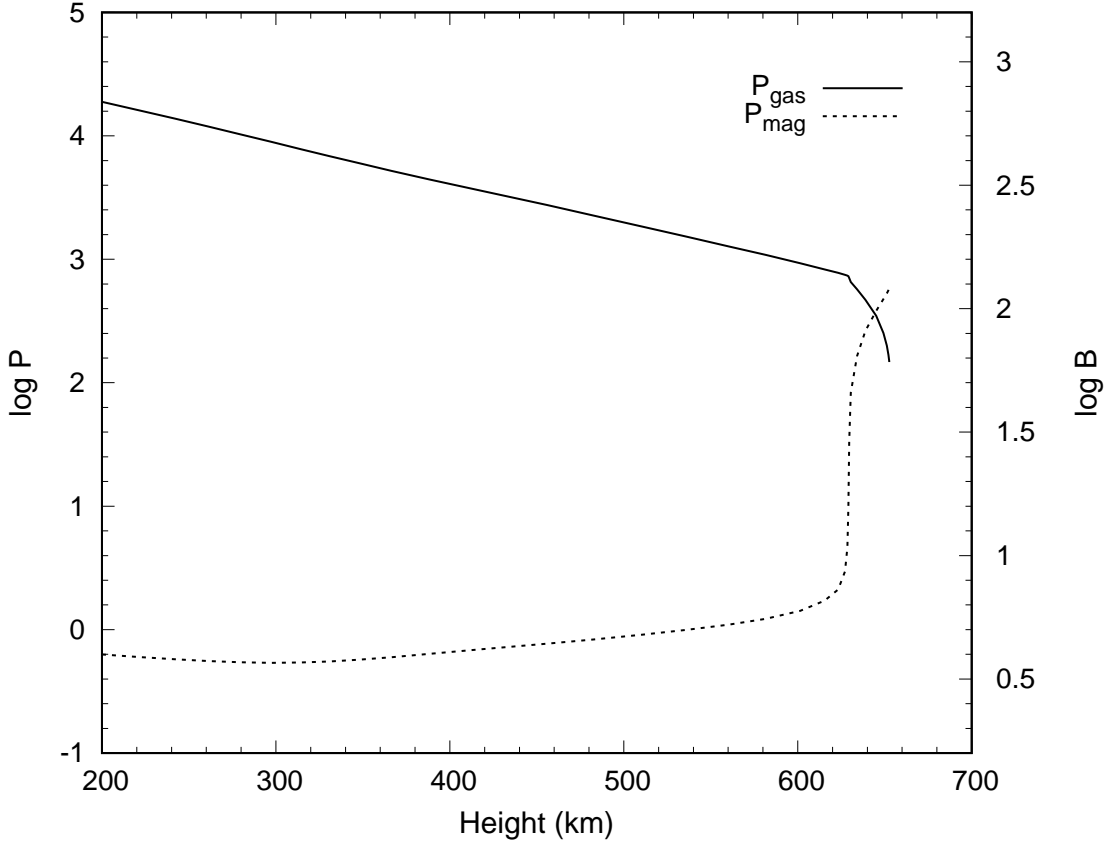}
\figurenum{4}
\caption{The distributions of the gas pressure and the magnetic pressure, as well as the strength of the magnetic field for the best-fit model by adopting the Hopf-grey atmosphere.}
\end{figure}

The distributions of the gas pressure and the magnetic pressure are shown respectively for the best-fit solar model in Figure 4. As pointed out by Rosenthal et al. (2002) and Cally (2007), the reflection and transmission of the p-mode oscillations occur at the gas pressure roughly equal to the magnetic pressure ($\beta ={{P}_{\text{gas}}}/{{P}_{\text{mag}}} \approx 1$). We use this condition to infer the strength of the magnetic field that is necessary for the p-mode oscillations to be reflected in the solar atmosphere. It can be found in Figure 4 that this condition corresponds to a magnetic field of about 90\,G. Such a magnetic field we have independently derived is in good agreement with the results of Trujillio Bueno et al. (2004) $60\sim 130$\,G, but considerably larger than that of Lites et al. (2008) 55\,G. As the amplitude of the measured polarization signals by the Zeeman effect depends sensitively on the cancellation of mixed magnetic polarities within the spatio-temporal resolution of the observations, our result is still consistent with that of Lites et al. (2008). The origin mechanism of the small-scale magnetic fields is still debating. A plausible interpretation is that the tangled magnetic fields are supplied with flux from the decaying flux tube fields. In particular, convection plays an important role in the expulsion of the existing magnetic field from the interior of a convective eddy flow like that of granules. On the other hand, small-scaled dynamos are the key process to maintain a turbulent magnetic field. However, their distribution in the photosphere is critical for the reflections of acoustic waves on these unsigned magnetic fields. Recent studies show that the power spectra of the internetwork magnetic fields agree well with the results obtained from local dynamo simulations (see, e.g., Buehler et al. 2013; Hotta et al. 2015; Danilovic et al. 2016). Our result on the strength of the tangled magnetic fields (about 90\,G) provides a crucial clue on the origin of the small-scale magnetic fields.

It can be found in Figure 4 that the magnetic-arch splicing layer is located at a height of about 630\,km in the photosphere. It is interesting to notice that a magnetic layer dominated by the horizontal fields has been revealed by the 3-D radiative MHD simulations. Steiner et al. (2008) found a local maximum of the horizontal field component at a height of around 500\,km , while Rempel (2014) found that the ratio of the horizontal to vertical field shows a maximum at a height of about 450\,km in the photosphere. The magnetic-arch splicing layer we have found has a similar location as those demonstrated by the above 3-D simulations. It should be noticed that the radiative MHD simulations use frequency-dependent opacity while we just adopt the Rosseland mean opacity from OPAL (Iglesias \& Rogers 1996) for the high temperature region and Ferguson et al.(2005) for the low temperature region. This difference may contribute to the uncertainties in the evaluation of the location of the magnetic-arch splicing layer. As pointed out by Galloway \& Weiss (1981) and Steiner et al. (2008), the magnetic field tends to be expelled by the convective motion not only in the lateral direction to the intergranular lanes but also in the vertical direction to the upper level of the photosphere. Therefore, the location of the magnetic-arch splicing layer directly corresponds to the accumulation of the horizontal magnetic flux that is expelled by the convective motions to the upper layer of the photosphere. In addition, the rapid increase of the magnetic pressure in Figure 4 corresponds closely to a configuration of magnetic arches with magnetic voids just in the below, which is often seen in the MHD simulations as ``small-scale canopies". And that is exactly why we refer to this magnetic configuration as the magnetic-arch splicing layer.

The heavy-element abundance is one of the fundamental ingredients to determine the structure and evolution of the sun (Basu 2016). The heavy elements increase the opacity, and thus significantly influence the internal structure of the sun. Besides, the heavy elements directly participate in the central hydrogen burning process through the CNO cycle, and therefore contribute to the solar neutrinos. We examine the matches of our best-fit models with different heavy-element abundance, and the result is shown in the left panel of Figure 5. It can be found that the solar model with the heavy-element abundance $(Z/X)_{\rm surf}$ around 0.0255 can best fit the p-mode frequencies of the sun, which is in agreement with the standard solar models (Basu 2016).

\begin{figure*}
\plotone{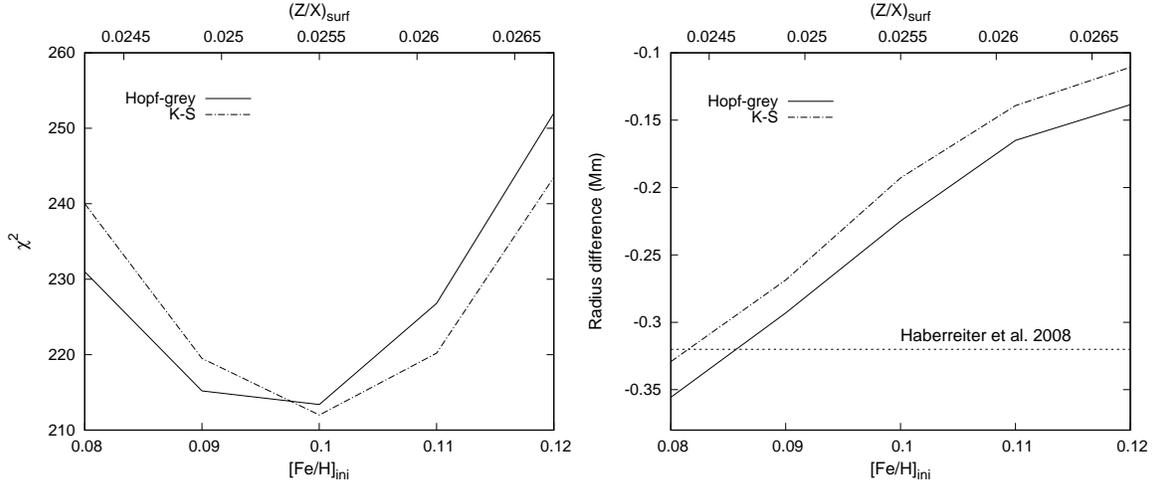}
\figurenum{5}
\caption{In the left panel the variations of $\chi^2$, which is defined as the sum of the squared differences of frequencies between observed and calculated ones weighted by the observational errors, are shown with the initial metallicity for the best-fit models by adopting different atmosphere models. In the right panel, the radius differences between the best-fit model and the sun are shown with the initial metallicity.}
\end{figure*}

As an important quantity that can be observed accurately, the solar radius is used to probe the static and dynamic structure of the solar interior and atmosphere (Kili\c c \& Golbasi 2011). Moreover, the solar diameter is also adopted in the climate model of the earth to get a realistic energy input reconstruction (Thuillier 2005). In the present study, we determine the best-fit solar models by matching the solar luminosity and the frequencies of the p-mode oscillations, leaving the stellar radius as a resulted parameter. In the right panel of Figure 5 we show the radius difference of our best-fit solar models with different heavy-element abundance to the sun. It can be found that all of our solar models have smaller radii than that of the sun with a few tenths in a megameter. Haberreiter et al. (2008) pointed out that this difference can be understood by the difference between the height at solar disk center where the optical depth $\tau_{\rm Ross}=2/3$ and the inflection point of the solar intensity profile on the limb, and they suggested that this difference is about 0.333Mm as shown in the right panel of Figure 5. However, they adopted two solar atmosphere models that do not include the magnetic field, and the seismic solar radius they adopted was inferred from only the observed f-mode frequencies (Schou et al. 1997). We include the magnetic field in the solar atmosphere and infer the seismic radius from all observed p-mode frequencies with the spherical harmonic degree ($l=0$, 1, 2, and 3), and our result is in good agreement with theirs.

\section{ Conclusions }

Small-scale magnetic fields are an important component in the solar photosphere. They contribute a significant amount of magnetic energy in the solar atmosphere, which can not only be responsible for the corona heating, but also influence the hydrostatic equilibrium in the photosphere itself. However, the distribution of the small-scale magnetic fields in the solar photosphere is not clear, i.e., randomly distributed or well-organized? In the present work, we identify that the small-scale canopies within the individual granules can act as a global magnetic-arch splicing layer, and investigate the effect of this magnetic-arch splicing layer on the propagations of the solar p-mode oscillations. Our main conclusions are summarized as follows.

1. The solar p-mode oscillations are totally reflected at the magnetic-arch splicing layer, resulting in tremendous improvement on the discrepancy between the observed and calculated p-mode frequencies.

2. The strength of the magnetic fields derived independently in our study is about 90\,G, and the location of the magnetic-arch splicing layer is 630\,km high in the upper photosphere. These results are quantitatively in agreement with the observations and 3-D MHD simulations.

3. The radii of the best-fit models are a few tenths in a megameter smaller than that of the sun, which is in agreement with Haberreiter et al. (2008). Our best-fit model has a metal abundance of Grevesse and Sauval (1998), which is in agreement with the standard solar models (Basu 2016).

\acknowledgments 
This work was supported by the National Natural Science Foundation of China (Nos. 11333006, 11973079, and 11521303). Fruitful discussions with Zhong Liu, Xiaoli Yan, and Mingde Ding are highly appreciated. We thank two anonymous reviewers for their productive comments and fruitful suggestions. 

\software{MESA(v11554; Paxton et al. 2011, 2013, 2015, 2018)}

{}

\clearpage 

\clearpage

\clearpage


\begin{thebibliography}{}

\bibitem[]{}Abbett, W.\,P. 2007, \apj, 665, 1469

\bibitem[]{}Aerts, C., Christensen-Dalsgaard, J., \& Kurtz, D.\,W. 2010, Asteroseismology, Astronomy and Astrophysics Library (Berlin: Springer)


\bibitem[]{}Bahcall J.\,N., Basu S., Pinsonneault, M., \& Serenelli, A.\,M. 2005, \apj, 618, 1049

\bibitem[]{}Ball, W.H. \& Gizon, L., 2014, \aap, 568, A123

\bibitem[]{}Ball, W.\,H., et al. 2016, \aap, 592, 159

\bibitem[]{}Basu, S. 2016, Living Rev. Sol. Phys., 13, 2

\bibitem[]{}Bogdan, T.\,J., et al. 2003, \apj, 599, 626

\bibitem[]{}Broomhall, A.-M., et al. 2009, \mnras, 396, L100

\bibitem[]{}Buehler D., Lagg A., \& Solanki, S.\,K. 2013, \aap, 555, 33

\bibitem[]{}Cally, P.\,S. 2007, Astron. Nachr., 328, 286

\bibitem[]{}Carlsson, M. \& Stein, R.F., 1994, in Proc. mini-workshop on chromospheric dynamics, Institute of theoretical astrophysics, Oslo, ed. M. Carlsson, p.47

\bibitem[]{}Christensen-Dalsgaard, J. 2008, \apss, 316, 113

\bibitem[]{}Christensen-Dalsgaard, J., D\"appen, W., \& Lebreton, Y., 1988,  Nature, 336, 634

\bibitem[]{}Christensen-Dalsgaard, J., et al. 1985, Nature, 315, 378

\bibitem[]{}Christensen-Dalsgaard, J., et al. 2010, ApJ, 713, L164

\bibitem[]{}Danilovic S., et al. 2010, \apj, 723, L149

\bibitem[]{}Danilovic S., et al. 2016, \aap, 594, 103

\bibitem[]{}Demarque, P., Guenther, D.B., \& Kim, Y.C., 1997, \apj, 474, 790

\bibitem[]{}Deubner, F.-L., 1975, \aap, 44, 371

\bibitem[]{}Dziembowski, W.A., Paterno, L., \& Ventura, R., 1988, \aap, 200, 213

\bibitem[]{}Ferguson, J.\,W., et al. 2005, \apj, 623, 585

\bibitem[]{}Finsterle, W., Jefferies, S.\,M., Cacciani, A., Rapex, P., \& McIntosh, S.\,W. 2004, \apj, 613, L185

\bibitem[]{}Galloway, D.\,J., \& Weiss, N.\,O. 1981, \apj, 243, 945

\bibitem[]{}Gough, D.\,O. 1990, in Lecture Notes in Physics Vol. 367, Progress of seismology of the sun and stars, ed. Y. Osaki, H. Shibahashi (Berlin: Springer), 281


\bibitem[]{}Grevesse, N., \& Sauval, A.\,J. 1998, Space Sci. Rev., 85, 161

\bibitem[]{}Grossmann-Doerth, U., Sch\"ussler, M., \& Steiner, O. 1998, \aap, 337, 928

\bibitem[]{}Haberreiter, M., Schmutz, W., \& Kosovichev, A.\,G. 2008, \apj, 675, L53

\bibitem[]{}Harvey, J.W., et al. 1996, Science, 272, 1284

\bibitem[]{}Hotta H., Rempel M., \& Yokoyama, T. 2015, \apj, 803, 42

\bibitem[]{}Houdek, G., Trampedach, R., Aarslev, M.\,J., \& Christensen-Dalsgaard, J. 2017, \mnras, 464, L124

\bibitem[]{}Iglesias, C.\,A., \& Rogers, F.\,J. 1996, \apj, 464, 943

\bibitem[]{}Keys, P.\,H., et al. 2020, \aap, 633, 60

\bibitem[]{}Kili\c c, H., \& Golbasi, O. 2011, \apss, 334, 75\\

\bibitem[]{}Kjeldsen H., Bedding T.\,R., \& Christensen-Dalsgaard, J. 2008, \apj, 683, L175

\bibitem[]{}Kosugi, T., et al. 2007, \solphys, 243, 3

\bibitem[]{}Krishna Swamy, K.\,S. 1966, \apj, 145, 174

\bibitem[]{}Lagg, A., et al. 2016, \aap, 596, 6

\bibitem[]{}Leenaarts, J., et al., 2007, \aap, 473, 625

\bibitem[]{}Leighton, R.B., Noyes, R.W., \& Simon, G.W., 1962, \apj, 135, 474

\bibitem[]{}Lites, B.\,W., et al. 2008, \apj, 672, 1237

\bibitem[]{}Mart\'inez Gonz\'alez, M.\,J., et al. 2016, \aap, 596, 5

\bibitem[]{}Morosin, R., et al. 2020, \aap, 642, 210

\bibitem[]{}Paxton, B., et al. 2011, \apjs, 192, 3

\bibitem[]{}Paxton, B., et al. 2013, \apjs, 208, 4

\bibitem[]{}Paxton, B., et al. 2015, \apjs, 220, 15

\bibitem[]{}Paxton, B., et al. 2018, \apjs, 234, 34

\bibitem[]{}Rempel, M. 2014, \apj, 789, 132

\bibitem[]{}Rogers, F.\,J., \& Nayfonov, A. 2002, \apj, 576, 1064

\bibitem[]{}Rosenthal, C.S., et al., 1999, \aap, 351, 689

\bibitem[]{}Rosenthal, C.\,S., et al. 2002, \apj, 564, 508

\bibitem[]{}Schaffenberger, W., Wedemeyer-B\"ohm, S., Steiner, O., \& Freytag, B. 2005, in ESA SP-596, Chromospheric and Coronal Magnetic Fields, ed. D.\,E. Innes, A. Lagg, \& S.\,K. Solanki (Katlenburg-Lindau: ESA), 65.1 

\bibitem[]{}Schaffenberger, W., Wedemeyer-B\"ohm, S., Steiner, O., \& Freytag, B. 2006, in ASP Conf. Ser. 354, Solar MHD: Theory and Observations, ed. J. Leibacher, R.\,F. Stein, \& H. Uitenbroek (San Francisco: ASP), 345

\bibitem[]{}Schou, J., \& Birch, A.\,C. 2020, \aap, 638, 51

\bibitem[]{}Schou, J., Kosovichev, A.\,G., Goode, P.\,R., \& Dziembowski, W.\,A. 1997, \apj, 489, L197

\bibitem[]{}Schrijver, C.\,J., \& Title, A. 2003, \apj, 597, L165

\bibitem[]{}Schrijver, C.\,J., et al. 1998, \nat, 394, 152

\bibitem[]{}Sch\"ussler, M., \& V\"ogler, A. 2008, \aap, 481, L5

\bibitem[]{}Sonoi, T., et al. 2019, \aap, 621, 84

\bibitem[]{}Steiner, O., et al. 2007, Astron. Nachr., 328, 323

\bibitem[]{}Steiner, O., Rezaei, R., Schaffenberger, W., \& Wedemeyer-B\"ohm, S. 2008, \apj, 680, L85

\bibitem[]{}Stenflo, J.\,O. 1987, \solphys, 114, 1

\bibitem[]{}Stenflo, J.\,O. 1994, Solar magnetic fields: polarized radiation diagnostics (Dordrecht: Kluwer)

\bibitem[]{}Stenflo, J.\,O. 2013, \aap, 555, 132

\bibitem[]{}Thoul, A.\,A., Bahcall, J.\,N., \& Loeb, A. 1994, \apj, 421, 828

\bibitem[]{}Thuillier, G., Sofia, S., \& Haberreiter, M. 2005, Adv. Space Res., 35, 329

\bibitem[]{}Trujillo Bueno, J., Shchukina, N., \& Asensio Ramos, A. 2004, \nat, 430, 326

\bibitem[]{}Unno, W., et al. 1989, Nonradial oscillations of stars, 2nd edn (Tokyo: Univ. of Tokyo Press)

\bibitem[]{}Vernazza, J.\,E., Avrett, E.H., \& Loeser, R. 1981, \apjs, 45, 635

\bibitem[]{}Wedemeyer, S., et al., 2004, \aap, 414, 1121

\bibitem[]{}Zeuner, F., et al. 2018, \aap, 619, 179

\bibitem[]{}Zeuner, F., et al. 2020, \apj, 893, L44


\end{thebibliography}
\end{document}